\documentclass{moriond}

\usepackage{amssymb}
\usepackage{amsmath}
\usepackage{amsfonts}
\usepackage{upgreek}
\usepackage{latexsym}
\usepackage{slashed,upgreek,amscd,cancel,tensor}
\usepackage{adjustbox}
\usepackage{epsfig}
\usepackage{cite}
\usepackage{hyperref}
\usepackage{amsfonts}
\usepackage{booktabs}
\usepackage{siunitx}
\usepackage{caption}
\usepackage{float}
\usepackage{color}

\def\be{\begin{equation}}
\def\ee{\end{equation}}
\def\beq{\begin{equation}}
\def\eeq{\end{equation}}
\def\bea{\begin{eqnarray}}
\def\eea{\end{eqnarray}}

\newcommand{\2}{{(2)}}
\newcommand{\3}{{(3)}}
\newcommand{\aq}{a}
\newcommand{\bb}{b}
\newcommand{\R}{{}^3\! R}
\newcommand{\aK}{\alpha_{\rm K}}
\newcommand{\aL}{\alpha_{\rm L}}
\newcommand{\aH}{\alpha_{\rm H}}
\newcommand{\aT}{\alpha_{\rm T}}
\newcommand{\aB}{\alpha_{\rm B}}
\newcommand{\bun}{\beta_1}
\newcommand{\bdeux}{\beta_2}
\newcommand{\btrois}{\beta_3}
\newcommand{\CI}{{\cal C}_{\rm I}}
\newcommand{\CII}{{\cal C}_{\rm II}}

\newcommand{\tg}{{\tilde g}}

\newcommand{\Photo}{}

\begin{document}
\vspace*{4cm}
\title{Degenerate Higher-Order Scalar-Tensor (DHOST) theories}

\author{David Langlois}

\address{ APC (Astroparticules et Cosmologie), CNRS-Universit\'e Paris Diderot,\\
 10, rue Alice Domon et L\'eonie Duquet, 75013 Paris, France}

\maketitle\abstracts{
This contribution  briefly reviews  scalar-tensor theories whose Lagrangian contains second-order derivatives of a scalar field but nevertheless propagate only one scalar mode (in addition to the usual two tensor modes), and are thus not plagued with the  Ostrodradsky instability. These theories, which  encompass the so-called Horndeski and Beyond Horndeski theories,  have recently been  fully classified  up to cubic order in second-order derivatives. After introducing these theories, I present a few phenomenological aspects. In cosmology, these theories can be included  in the unified effective description of dark energy and modified gravity. Finally, neutron star solutions in some specific models are discussed.}

\section{Introduction}
There have been numerous attempts to modify or extend general relativity, with either the motivation to account for  dark energy (and sometimes dark matter) or, more modestly, to construct benchmark models that are useful to  test general relativity quantitatively. 
Scalar-tensor theories have often played a prominent role in these attempts  and, lately, special attention has been devoted to scalar-tensor theories whose Lagrangians contain second-order derivatives of a scalar field. 

Lagrangians of this type, which contain ``accelerations'', are generically plagued by an instability due to the presence, in addition to the usual scalar mode and  tensor modes,  of an extra scalar degree of freedom (unless the higher order terms can be treated as perturbative terms in the sense of low energy effective theories).   Until recently, it was believed that only theories that yield second-order Euler-Lagrange equations were free of this dangerous extra degree of freedom.
In the last couple of years, it has been realized that there in fact exists a much larger class of theories that satisfy this property. 

\section{From Horndeski to DHOST theories}
\label{section_DHOST}

\subsection{Higher-Order Scalar-Tensor theories}
In this section, we introduce  scalar-tensor theories whose action is a functional of  a metric $g_{\mu\nu}$ and of a scalar field $\phi$, allowing  for a dependence not only on $\phi$ and its gradient $\phi_\mu\equiv\nabla_\mu\phi$ as  usual,  but also on its second derivatives $\phi_{\mu\nu}\equiv\nabla_\mu\!\nabla_\nu\phi$.
 Restricting our investigation to Lagrangians that depend on  $\phi_{\mu\nu}$ { up to cubic order}, we are interested by actions of the form
 \bea
\label{action}
S[g,\phi] &=& \int d^4 x \, \sqrt{- g }
\left[ f_0(X,\phi) + f_1(X,\phi) \Box \phi
+
f_2(X,\phi) \, R+ C_\2^{\mu\nu\rho\sigma} \,  \phi_{\mu\nu} \, \phi_{\rho\sigma}+
\right.
\cr
&&
\left. \qquad \qquad\qquad + f_3(X, \phi) \, G_{\mu\nu} \phi^{\mu\nu}  +  
C_\3^{\mu\nu\rho\sigma\alpha\beta} \, \phi_{\mu\nu} \, \phi_{\rho\sigma} \, \phi_{\alpha \beta} \right]  \;,
\eea
where the functions $f_i$   depend only on $\phi$ and $X \equiv \phi_\mu \phi^\mu$, while  $R$ and $G_{\mu\nu}$ denote, respectively, the usual Ricci scalar and Einstein tensor associated with the four-dimensional metric $g_{\mu\nu}$. 
The tensors 
$\mathbf{C}_\2$ and $\mathbf{C}_\3$ are the most
general tensors constructed with the metric $g_{\mu\nu}$ and  the scalar field gradient  $\phi_\mu$.

It is easy to see that the terms quadratic in $\phi_{\mu\nu}$  can be rewritten as
\beq
\label{C2}
 C_\2^{\mu\nu\rho\sigma} \,  \phi_{\mu\nu} \, \phi_{\rho\sigma} =\sum_{A=1}^{5}a_A(X,\phi)\,   L^\2_ A\,,
\eeq
with 
\be
\label{QuadraticL}
\begin{split}
& L^\2_1 = \phi_{\mu \nu} \phi^{\mu \nu} \,, \qquad
L^\2_2 =(\Box \phi)^2 \,, \qquad
L_3^\2 = (\Box \phi) \phi^{\mu} \phi_{\mu \nu} \phi^{\nu} \,,  \\
& L^\2_4 =\phi^{\mu} \phi_{\mu \rho} \phi^{\rho \nu} \phi_{\nu} \,, \qquad
L^\2_5= (\phi^{\mu} \phi_{\mu \nu} \phi^{\nu})^2\,,
\end{split}
\ee
where the $a_A$ are five arbitrary functions of $X$ and $\phi$.
Similarly, the cubic terms can be written in terms of ten arbitrary functions $b_A$, as
\beq
\label{C3}
C_\3^{\mu\nu\rho\sigma\alpha\beta} \, \phi_{\mu\nu} \, \phi_{\rho\sigma} \, \phi_{\alpha \beta} = \sum_{A=1}^{10} b_A(X,\phi)\,  L^\3_A \,,
\eeq
where 
\be
\label{CubicL}
\begin{split}
& L^\3_1=  (\Box \phi)^3  \,, \quad
L^\3_2 =  (\Box \phi)\, \phi_{\mu \nu} \phi^{\mu \nu} \,, \quad
L^\3_3= \phi_{\mu \nu}\phi^{\nu \rho} \phi^{\mu}_{\rho} \,,   \quad
 L^\3_4= \left(\Box \phi\right)^2 \phi_{\mu} \phi^{\mu \nu} \phi_{\nu} \,, \\
&L^\3_5 =  \Box \phi\, \phi_{\mu}  \phi^{\mu \nu} \phi_{\nu \rho} \phi^{\rho} \,, \quad
L^\3_6 = \phi_{\mu \nu} \phi^{\mu \nu} \phi_{\rho} \phi^{\rho \sigma} \phi_{\sigma} \,,   \quad
 L^\3_7 = \phi_{\mu} \phi^{\mu \nu} \phi_{\nu \rho} \phi^{\rho \sigma} \phi_{\sigma} \,, \\
&L^\3_8 = \phi_{\mu}  \phi^{\mu \nu} \phi_{\nu \rho} \phi^{\rho}\, \phi_{\sigma} \phi^{\sigma\lambda} \phi_{\lambda} \,,   \quad
 L^\3_9 = \Box \phi \left(\phi_{\mu} \phi^{\mu \nu} \phi_{\nu}\right)^2  \,, \quad
L^\3_{10} = \left(\phi_{\mu} \phi^{\mu \nu} \phi_{\nu}\right)^3 \,.
\end{split}
\ee

Theories described by  an  action of the form  (\ref{action})  in general  contain, in addition to the usual scalar mode and two tensor modes, an extra scalar mode  leading to the so-called  Ostrogradsky instability \cite{Woodard:2015zca}. 
 However, by imposing some restrictions  on the  functions $f_2$,  $f_3$ and $\aq_A$ and $\bb_A$, it is possible to find
 theories that contain only one propagating scalar mode.
 
Historically, theories of this type were found  in several steps. The starting point was  the construction of higher-order theories leading to at most second-order Euler-Lagrange equations for the metric and for the scalar field, initially due to Horndeski~\cite{Horndeski:1974wa}~\footnote{The work of Horndeski had been completely forgotten until it was resurrected in \cite{Charmousis:2011bf}. At that time, the same theories had just been rediscovered, under the name of generalised Galileons, in \cite{Deffayet:2011gz}. Their equivalence was in particular shown in the Appendix of \cite{Kobayashi:2011nu} (v3 on arXiv).}.  Until recently, it was (wrongly) believed that requiring at most second-order Euler-Lagrange equations  was necessary  to get rid of the extra scalar degree of freedom and, as a consequence,  Horndeski theories were considered to be the most general theories without Ostrogradsky instability. This belief was  challenged by  a new class of  theories, now often called Beyond Horndeski (or GLPV),  proposed in \cite{Gleyzes:2014dya,Gleyzes:2014qga}, extending Horndeski's theories and leading to higher-order equations of motion~\footnote{It is also worth noting that  \cite{Zumalacarregui:2013pma} had already pointed out the possibility to   construct  theories  ``beyond Horndeski" by applying disformal transformations of the metric to the Einstein-Hilbert action.}.  Beyond Horndeski theories were finally superseded by a larger class of theories, the DHOST theories, once it was understood that the crucial element that characterizes higher-order theories propagating a single scalar degree of freedom is the degeneracy of their Lagrangian \cite{Langlois:2015cwa,Langlois:2015skt}, rather than the order of their equations of motion.
By using the degeneracy criterium, the quadratic DHOST theories were first identified in \cite{Langlois:2015cwa}, extending both quadratic Horndeski and Beyond Horndeski Lagrangians~\footnote{The name DHOST was not coined in the original paper \cite{Langlois:2015cwa} but later in \cite{Achour:2016rkg}. Note that the very same (quadratic) theories were also dubbed  ``Extended Scalar-Tensor" in \cite{Crisostomi:2016czh}.}. More recently, all DHOST theories up to cubic order have been systematically classified~\cite{BenAchour:2016fzp}.

\subsection{Degeneracy of the Lagrangian}
As mentioned above, the crucial ingredient that singles out  higher-order theories with a single scalar degree of freedom  is the degeneracy of their Lagrangian~\cite{Langlois:2015cwa,Langlois:2015skt}.
To better  understand this notion of degeneracy, it is instructive to present a very simple toy model based on classical point dynamics. Let us consider the Lagrangian
\beq
\label{L0}
L=\frac12 a\, \ddot\phi^2+b\,\ddot\phi \,\dot q+\frac12 \dot\phi^2+\frac12 c\, \dot q^2-V(\phi, q)\,,
\eeq
where $a$, $b$ and $c$ are constant coefficients and $V(\phi, q)$ is some potential. This Lagrangian involves the acceleration of $\phi$ 
but not that of $q$. If $a\neq 0$,  one gets fourth-order equations of motion, whereas, if $a=0$ but $b\neq 0$, one obtains third-order equations of motion. 

In order to work with a more familiar Lagrangian containing only velocities, let us introduce, following \cite{Langlois:2015cwa},  the auxiliary variable 
\beq
Q\equiv \dot \phi\,,
\eeq
leading to the new (and  equivalent) Lagrangian
\beq
L=\frac12 a\, \dot Q^2+b\,\dot Q \,\dot q+\frac12 c\,  \dot q^2 +\frac12 Q^2-V(\phi, q) -\lambda (Q-\dot \phi)\,,
\eeq
which does not include any acceleration. 
 
Let us now try to identify the number of independent physical degrees of freedom in the system. Equivalently, one can count the number of initial conditions that are needed to fully determine the system at some initial time. From the equations of motion, it is easy to see that two cases arise, depending on the nature of the Hessian matrix, defined by
\beq
M\equiv \left(\frac{\partial^2L}{\partial v^a\partial v^b}\right)=
\left(
\begin{array}{cc}
a & b \cr
b & c
\end{array}
\right)\,,
\eeq
where the symbol $v^a$ denotes the velocities, i.e. $v^a\equiv \{\dot Q, \dot q\}$. 

In the generic case where  $M$ is invertible,  one finds that six initial conditions are needed, which corresponds to the existence of three degrees of freedom. While the variable $q$ describes as usual one degree of freedom, the variable $\phi$  is associated with two degrees of freedom. 
By contrast, in the particular cases where $M$ is degenerate, i.e. 
\beq
\det M=ac-b^2 =0\,,
\eeq
only four initial conditions are necessary, which means that only two degrees of freedom are present. The extra mode associated with $\phi$ is eliminated when $M$ is degenerate. By extension, it can be said that the initial Lagrangian (\ref{L0}) is degenerate in this situation.

The number of degrees of freedom can also be determined by resorting to a Hamiltonian analysis. When the Lagrangian is degenerate,   the conjugate momenta satisfy a (primary) constraint. By writing down the time evolution of this constraint, one finds that it leads to a secondary constraint in phase space. These two constraints eliminate one degree of freedom, in agreement with the analysis based on  the equations of motion.

The above discussion can be generalised to the case of $n$ variables similar to $\phi$. In order to get rid of all of the $n$ extra degrees of freedom that arise in general,   one must not only impose a degeneracy of order $n$ of the Hessian matrix, which guarantees the existence of $n$ primary constraints, but also require additional constraints to ensure the presence of $n$ secondary constraints~\cite{Motohashi:2016ftl,Klein:2016aiq}.

\subsection{Horndeski and Beyond Horndeski theories}
Well-known particular examples of DHOST theories are  Horndeski's  theories. 
They are characterized by four arbitrary functions of $\phi$ and $X$, corresponding to the four functions $f_A$ that appear in the general action (\ref{action}). The other functions $a_A$ and $b_A$ are then completely determined in terms of $f_2$ and $f_3$, respectively. 
The quadratic part of the Horndeski action, which it is convenient to  denote  $L^{\rm H}_\2[f_2]$, is thus fully determined by the function $f_2$, with   the quadratic coefficient $\aq_A$  given by
\beq
\aq_1=-\aq_2=2 f_{2,X}\,, \qquad \aq_3=\aq_4=\aq_5=0\,.
\eeq
Similarly, the cubic part of Horndeski theories, $L^{\rm H}_\3[f_3]$,  depends only on the functions $f_3$, while
\beq
3\bb_1=-\bb_2=\frac32 \bb_3=f_{3,X}\,, \qquad \bb_A=0 \ (A=4,\dots, 10)\,.
\eeq

The so-called Beyond Horndeski (or GLPV) theories, introduced in \cite{Gleyzes:2014dya,Gleyzes:2014qga}, extend Horndeski theories by including  two additional Lagrangians, each characterized by a single arbitrary function. The first of these  Lagrangians, which can be written as $L_\2^{\rm bH}[g_2]$,  is quadratic and characterized by the coefficients
\beq
\aq_1=-\aq_2=X g_2\,, \qquad \aq_3=-\aq_4= 2g_2\,, \qquad \aq_5=0\,.
\eeq
The second new Lagrangian, which is cubic and will be denoted $L_\3^{\rm bH}[g_3]$, depends on a single arbitrary function $g_3$ and its  non vanishing coefficients $b_A$ are given by
\begin{equation}
\frac{b_1}{X} = - \frac{b_2}{3X}  = \frac{b_3}{2X} = - \frac{b_4}{3} = \frac{b_5}{6} = \frac{b_6}{3} = - \frac{b_7}{6} = g_3 \,. 
\end{equation}

In the original paper \cite{Gleyzes:2014dya}, it was not yet fully clear whether arbitrary sums of the four Horndeski Lagrangians and of the two new Lagrangians $L_\2^{\rm bH}[g_2]$ and $L_\3^{\rm bH}[g_3]$ contained a single scalar mode.  It was first pointed out   in \cite{Gleyzes:2014qga} that some Beyond Horndeski theories  could be  related to Horndeski theories via (invertible) conformal-disformal transformations, in which case they  should have the same number of degrees of freedom as their Horndeski counterparts~\footnote{Note that the calculation in the final part of \cite{Gleyzes:2014qga}, directly inspired  by a similar calculation  in\cite{Zumalacarregui:2013pma}, does not lead to a manifestly second-order system, as originally claimed. But the main point of the paper, based on disformal transformations, remains valid.}. But it was only with the concept of degeneracy that this question was finally settled, with the results of  \cite{Langlois:2015cwa,Langlois:2015skt} and \cite{Crisostomi:2016tcp}. 

Let us briefly  summarize these results, by stressing that the sum of two degenerate Lagrangians is not necessarily degenerate.  Moreover the terms $f_0$ and $f_1 \Box\phi$ can always be added in the action without modifying the degeneracy of the total Lagrangian, so we do not need to worry about these terms any further. For the remaining terms, the following combinations involving Beyond Horndeski terms are degenerate:
$L_\2^{\rm H}+L_\2^{\rm bH}$, $L_\3^{\rm H}+L_\3^{\rm bH}$ and  $L_\2^{\rm bH}+L_\3^{\rm bH}$. By contrast, the following combinations are not degenerate: $L_\2^{\rm H}+L_\2^{\rm bH}+L_\3^{\rm H}$, $L_\2^{\rm H}+L_\2^{\rm bH}+L_\3^{\rm bH}$.

\subsection{DHOST theories}
As discussed above, the crucial element that characterizes higher-order theories with a single scalar degree of freedom is the degeneracy of their Lagrangian, hence their name DHOST~\footnote{Amusingly,  this acronym can be obtained by substituting the initial of 'ghost' with  'd' of 'degeneracy'.}. 

 DHOST theories were originally identified  at  quadratic order in $\phi_{\mu\nu}$
  (i.e.~with the functions $f_2$ and $\aq_A$ only)    in \cite{Langlois:2015cwa} and a complete Hamiltonian analysis in \cite{Langlois:2015skt} soon confirmed that they indeed contained only one scalar degree of freedom. Quadratic DHOST theories were  further studied in \cite{Crisostomi:2016czh,Achour:2016rkg,deRham:2016wji}. More recently, the  identification of DHOST theories has  been extended up to cubic order, i.e.~by including the second line of (\ref{action}), in \cite{BenAchour:2016fzp} and the interested reader will find the full classification there.

In summary, DHOST theories include  seven  subclasses of quadratic theories (four classes with $f_2\neq 0$ and three classes with $f_2=0$) and  nine subclasses of  cubic theories (two with $f_3\neq 0$ and seven with $f_3=0$). These quadratic and cubic subclasses can be combined to yield degenerate hybrid theories, involving both quadratic and cubic terms, but all combinations are not possible: only 25 combinations (out of 63)  lead to degenerate theories, 
often with extra conditions on the functions $a_A$ and $b_A$ in the Lagrangian (see \cite{BenAchour:2016fzp} for  details and for the explicit form of the functions in each subclass).

\subsection{Disformal transformations}
A legitimate  question about this classification is whether seemingly different DHOST theories could correspond the same theory in  different guises, in other words whether some theories could be identified via field redefinitions~\footnote{The coupling to matter is ignored here. If, after a redefinition of the metric, two related theories are minimally coupled to matter, then they are physically distinct.}.
Since the Lagrangian depends on a metric and on a scalar field, natural field redefinitions of the metric involve disformal transformations~\cite{Bekenstein:1992pj}
\beq
\label{disformal}
\tg_{\mu\nu}=C(X, \phi) g_{\mu\nu}+D(X, \phi) \, \phi_\mu\, \phi_\nu\,.
\eeq
Via this transformation,  any action $\tilde S$ given as a functional  of  $\tg_{\mu\nu}$ and $\phi$ induces a new   action $S$  for  $g_{\mu\nu}$ and $\phi$, when one substitutes the above expression for  $\tg_{\mu\nu}$ in $\tilde S$: 
 \beq
S[\phi, g_{\mu\nu}]\equiv\tilde S\left[\phi, \tg_{\mu\nu}=C \,g_{\mu\nu}+D \, \phi_\mu\phi_\nu\right]\,.
\eeq
The actions $S$ and $\tilde{S}$ are then said to be related by the disformal transformation (\ref{disformal}).
The disformal transformations of all quadratic DHOST theories have been investigated in \cite{Achour:2016rkg}, where it was shown that all seven subclasses are stable under  the action of disformal transformations. 

Interestingly, there is a nice correspondence between the type of  disformal transformations and the extent of the corresponding stable class of theories:
\begin{itemize}
\item Horndeski theories are stable under disformal transformations characterized by $C(\phi)$ and $D(\phi)$, i.e.  conformal and disformal factors that depend only on $\phi$, but not on $X$~\cite{Bettoni:2013diz}.

\item Beyond Horndeski theories are stable under disformal transformations characterized by $C(\phi)$ and $D(\phi, X)$~\cite{Gleyzes:2014qga}.

\item Finally, DHOST theories are stable under the most general disformal transformations where $C$ and $D$ depend on both $\phi$ and $X$~\cite{Achour:2016rkg}.

\end{itemize}

\section{Cosmology and astrophysics}
After the short introduction  to DHOST theories given in the previous section, let us now discuss briefly some phenomenological consequences of these theories in the context of cosmology and of astrophysics. 

\subsection{Cosmology}
In order to study the cosmology of DHOST theories, it is very convenient to resort to the unified formalism that has been developed for an  effective description  of Dark Energy and Modified Gravity (see e.g. \cite{Gleyzes:2014rba} for a review). 

This approach is based on  a $3+1$  decomposition of spacetime,  in which the spatial slices coincide with uniform scalar field hypersurfaces. In this particular gauge, sometimes called unitary gauge, the action of DHOST theories is of the form 
\beq
S= \int d^3x \,  dt \,  N\sqrt{h}\,  L[N, K_{ij}, \R_{ij};t]\,,
\eeq
where $N$ is the lapse function [which appears in the $3+1$ form of the spacetime metric $ds^2=-N^2 dt^2 +h_{ij} (dx^i+N^i dt)(dx^j+N^jdt)$, $N^i$ being the shift vector, $h_{ij}$ the spatial metric]; $K_{ij}$ is the extrinsic curvature tensor and $\R_{ij}$ the intrinsic curvature tensor. 

The Friedmann equations associated with  a spatially   flat Friedmann-Lema\^itre-Robertson-Walker (FLRW) spacetime $ds^2 = - \bar{N}^2(t) dt^2 + a^2(t) \delta_{ij} dx^i dx^j$, are then simply  derived from the homogeneous action
\beq
S_{\rm homog}= \int   dt \,  N a^3 L[N=\bar{N}(t), K^i_j=\frac{\dot a}{\bar{N}a}\delta^i_j, \R_{ij}=0;t]\,.
\eeq
To study the dynamics of linear perturbations, one needs to write down the action at quadratic order in perturbations. These perturbations are associated with the three basic ingredients of the action:
\beq
\delta N\equiv N-\bar{N}\,, \qquad \delta K^i_j=K^i_j-H\delta^i\,, \qquad \delta\, \R^i_j=\R^i_j\,,
\eeq
where $H=\dot a/\bar{N}a$ is the Hubble parameter, and $\R^i_j$ is already a perturbation since it vanishes in the background.
The Lagrangian at quadratic order is then obtained via a Taylor expansion, which is formally written as
\beq
L(q_A)=\bar{L}+\frac{\partial L}{\partial q_A}\delta q^A+\frac12 \frac{\partial^2 L}{\partial q_A\partial q_B}\delta q^A \delta q^B+\dots \,.
\eeq
where $q^A=\{N, K^i_j, \R^i_j\}$.

All (quadratic and cubic) DHOST theories lead to a Lagrangian quadratic in linear perturbations of the form~\cite{Langlois:2017mxy}
\begin{eqnarray}
\label{Squad}
 S_{\rm quad} &=& \int d^3x \,  dt \,  a^3  \frac{M^2}2\bigg\{ \delta K_{ij }\delta K^{ij}- \left(1+\frac23\aL\right)\delta K^2  +(1+\aT) \bigg( \R \frac{\delta \sqrt{h}}{a^3} + \delta_2 \R 
 \bigg)
  + H^2\aK \delta N^2
 \cr
&&  +4 H \aB \delta K \delta N+ ({1+\aH}) \, \R  \, \delta N   +  4 \bun  \delta K  {\delta \dot N }   + \bdeux  {\delta \dot N}^2 +  \frac{\btrois}{a^2}(\partial_i \delta N )^2   
\bigg\} \; ,
\end{eqnarray}
where $\delta_2\R$ denotes the second order term in the perturbative expansion of $\R$, where the parameters $M$, $\aL$, $\aT$, $\aK$, $\aB$, $\aH$, $\bun$, $\bdeux$ and $\btrois$ are time-dependent functions. Moreover, one finds that these parameters, for DHOST theories, are restricted to    satisfy either one of the following sets of conditions:
\beq
\label{Ia}
\CI:\qquad \aL=0\,, \qquad \bdeux=-6\bun^2\,,\qquad   \btrois=-2\bun\left[2(1+\aH)+\bun (1+\aT)\right]\,,
\eeq
or 
\beq
\label{IIa}
\CII:\qquad \bun=- (1+\aL)\frac{1+\aH}{1+\aT}\,, \quad \bdeux=-6(1+\aL) \frac{(1+\aH)^2 }{(1+\aT)^2}\,,\quad \beta_3=2\frac{(1+\aH)^2}{1+\aT}\,.
\eeq
The category $\CI$ contains the subclass of Horndeski theories and of those related to Horndeski via disformal transformations. 

From the action (\ref{Squad}), one can isolate the physical degrees of freedom, which reduce to one scalar and two tensor modes for DHOST. Their action is given by~\cite{Langlois:2017mxy}
\beq
 S_{\rm quad, phys} = \int d^3x \,  dt \,  a^3  \bigg\{\frac{M^2}2\left[ A\,\dot{\tilde\zeta}^2-B\, \frac{(\partial{\tilde\zeta})^2}{a^2}\right]+\frac{M^2}8\left[\dot\gamma_{ij}^2-\frac{1+\aT}{a^2}(\partial_k\gamma_{ij})^2\right]\bigg\} \,,
\eeq
where $\tilde\zeta\equiv \zeta-\bun \delta N$, $\zeta$ being the usual curvature perturbation of the spatial part of the metric and $h_{ij}$ denotes the transverse-traceless perturbation of the  metric. The explicit expressions for the coefficients $A$ and $B$ can be found in \cite{Langlois:2017mxy}. In particular, for models in the category $\CII$, one finds that $B=-(1+\aT)$. Comparing with the tensor part, one sees that the coefficients of the gradient terms for the scalar and tensor modes have opposite signs and therefore, these modes cannot be stable simultaneously. This signals an instability for theories satisfying $\CII$.

Another problem for $\CII$  theories is the divergence of  the effective Newton's constant, 
\beq
8\pi G_{\rm N}=M^{-2}\left[\frac{(1+\aH)^2}{1+\aT}-\frac{\btrois}{2}\right]^{-1}\,,
\eeq
defined in the static linear regime around Minkowski~\cite{Langlois:2017mxy}.

Finally, one should also include the description of matter.
The coupling of matter to the metric  can be either minimal or nonminimal. At the linear level, it is easy compute how the parameters that describe the matter coupling change  under disformal transformations (\ref{disformal}). Similar transformations exist for the parameters of (\ref{Squad}). See \cite{Langlois:2017mxy} for details. 

\subsection{Stars in Beyond Horndeski theories}
Even if the main motivation for  modified gravity arises from the observed acceleration of the cosmological expansion, it is indispensable to verify that any viable theory  remains compatible with astrophysical observations and solar system constraints. 

As part of this programme, let us concentrate on Beyond Horndeski theories. In these models,  it has been noticed that the Vainshtein mechanism is partially broken inside matter~\cite{Kobayashi:2014ida}. For spherical bodies, a new term appears in the gravitational law,
\beq
\label{Upsilon}
\frac{d\Phi}{dr}=\frac{G_{\rm N} M(r)}{r^2}+\frac{\Upsilon }{4} G_{\rm N}\frac{d^2M(r)}{dr^2}\,,
\eeq
where $\Phi$ is the gravitational potential and $M(r)$ is the mass inside a sphere of radius $r$. This leads to a modified profile of Newtonian stars~\cite{Koyama:2015oma,Saito:2015fza}.

Following the recent works\cite{Babichev:2016jom} and \cite{Sakstein:2016oel}, let us  discuss  neutron stars in a specific model
described by the action
\begin{equation}
\label{eq:action}
S=\int d^4x\,  \sqrt{-g}\, \left[
M_{\rm P}^2\left(\frac{R}{2}-k_0\Lambda\right)-k_2X -\frac{\zeta}{2} X^2+\mathcal{L}_{\2,{\rm bH}}[g_2]\right]\,,
\end{equation}
where  $k_0$, $k_2$, $\zeta$ and $g_2$  are assumed to be constant. 
By writing the Friedmann equations for this model, one can easily find de Sitter  solutions where the Hubble parameter $H$ is constant.
In order to embed a spherical object within such cosmological spacetime, it is useful to rewrite the de Sitter solution  in Schwarzschild-like coordinates, 
\begin{eqnarray}
\label{dS}
ds^2 &=&-(1-H^2 r^2) dt^2+\frac{dr^2}{1-H^2 r^2}+r^2\left(d\theta^2+\sin^2\theta d\phi^2\right)\,,
\\
\phi(r,t)&=&v_0 t+\frac{v_0}{2H} \ln (1-H^2 r^2)\,.
\end{eqnarray}
One can then insert a spherical symmetric object in this cosmological solution by trying to solve the Einstein equations for a metric of the form  
\beq
ds^2=-e^{\nu(r)}dt^2+e^{\lambda(r)}dr^2+r^2\left(d\theta^2+\sin^2\theta d\phi^2\right)\,,
\eeq
going asymptotically to (\ref{dS}). 
The energy-momentum tensor receives a contribution from the scalar field as well as a contribution from a perfect fluid which is assumed to model the neutron star's matter. Remarkably, one can find an exact solution outside the star, corresponding to a Schwarzschild-de Sitter geometry. Inside the star, the equations of motion can be solved  numerically, assuming some equation of state, in order to determine the matter density profile and the internal geometry. 

When $\Upsilon <0$, one finds that stars with fixed mass have a larger radius than their GR counterparts. Moreover, for the same equation of state, the maximum mass can increase significantly with respect to GR\cite{Babichev:2016jom}. Modified gravity could thus provide a solution to the hyperon puzzle. Moreover, one can derive a relation between the dimensionless moment of inertia $Ic^2/G^2M^3$ and the compactness $GM/Rc^2$, which is robust in the sense that it weakly depends on the equation of state and which can discriminate between modified gravity and GR\cite{Sakstein:2016oel}.

\section*{Acknowledgments}
I would like to thank my numerous collaborators, especially Karim Noui and Filippo Vernizzi, for their contributions to the works presented here.


\section*{References}

\begin{thebibliography}{99}

\bibitem{Woodard:2015zca} 
  R.~P.~Woodard,
  Scholarpedia {\bf 10}, no. 8, 32243 (2015)
  [arXiv:1506.02210 [hep-th]].

\bibitem{Horndeski:1974wa} 
  G.~W.~Horndeski,
  Int.\ J.\ Theor.\ Phys.\  {\bf 10}, 363 (1974).
  
  
\bibitem{Charmousis:2011bf} 
  C.~Charmousis, E.~J.~Copeland, A.~Padilla and P.~M.~Saffin,
  Phys.\ Rev.\ Lett.\  {\bf 108}, 051101 (2012)
  [arXiv:1106.2000 [hep-th]].

\bibitem{Deffayet:2011gz} 
  C.~Deffayet, X.~Gao, D.~A.~Steer and G.~Zahariade,
  Phys.\ Rev.\ D {\bf 84}, 064039 (2011)
  [arXiv:1103.3260 [hep-th]].


\bibitem{Kobayashi:2011nu} 
  T.~Kobayashi, M.~Yamaguchi and J.~Yokoyama,
  Prog.\ Theor.\ Phys.\  {\bf 126}, 511 (2011)
  [arXiv:1105.5723 [hep-th]].

  
\bibitem{Gleyzes:2014dya} 
  J.~Gleyzes, D.~Langlois, F.~Piazza and F.~Vernizzi,
  Phys.\ Rev.\ Lett.\  {\bf 114}, no. 21, 211101 (2015)
  [arXiv:1404.6495 [hep-th]].
  
\bibitem{Gleyzes:2014qga} 
  J.~Gleyzes, D.~Langlois, F.~Piazza and F.~Vernizzi,
  JCAP {\bf 1502}, 018 (2015)
  [arXiv:1408.1952 [astro-ph.CO]].

  
\bibitem{Zumalacarregui:2013pma} 
  M.~Zumalacarregui and J.~Garcia-Bellido,
  Phys.\ Rev.\ D {\bf 89}, 064046 (2014)
  [arXiv:1308.4685 [gr-qc]].
  


\bibitem{Langlois:2015cwa} 
  D.~Langlois and K.~Noui,
  JCAP {\bf 1602}, no. 02, 034 (2016)
  [arXiv:1510.06930 [gr-qc]].
  
\bibitem{Langlois:2015skt} 
  D.~Langlois and K.~Noui,
  JCAP {\bf 1607}, no. 07, 016 (2016)
  [arXiv:1512.06820 [gr-qc]].

  
\bibitem{Achour:2016rkg} 
  J.~Ben Achour, D.~Langlois and K.~Noui,
  Phys.\ Rev.\ D {\bf 93}, no. 12, 124005 (2016)
  [arXiv:1602.08398 [gr-qc]].

\bibitem{Crisostomi:2016czh} 
  M.~Crisostomi, K.~Koyama and G.~Tasinato,
  JCAP {\bf 1604}, no. 04, 044 (2016)
  [arXiv:1602.03119 [hep-th]].
 
\bibitem{BenAchour:2016fzp} 
  J.~Ben Achour, M.~Crisostomi, K.~Koyama, D.~Langlois, K.~Noui and G.~Tasinato,
  JHEP {\bf 1612}, 100 (2016)
  [arXiv:1608.08135 [hep-th]].

 
  
    
\bibitem{Motohashi:2016ftl} 
  H.~Motohashi, K.~Noui, T.~Suyama, M.~Yamaguchi and D.~Langlois,
  JCAP {\bf 1607}, 007 (2016)
  [arXiv:1603.09355 [hep-th]].
  
  
\bibitem{Klein:2016aiq} 
  R.~Klein and D.~Roest,
  JHEP {\bf 1607}, 130 (2016)
  [arXiv:1604.01719 [hep-th]].
  
 
  
\bibitem{Crisostomi:2016tcp} 
  M.~Crisostomi, M.~Hull, K.~Koyama and G.~Tasinato,
  JCAP {\bf 1603}, no. 03, 038 (2016)
  [arXiv:1601.04658 [hep-th]].


    
\bibitem{deRham:2016wji} 
  C.~de Rham and A.~Matas,
  JCAP {\bf 1606}, no. 06, 041 (2016)
  [arXiv:1604.08638 [hep-th]].
  

\bibitem{Bekenstein:1992pj} 
  J.~D.~Bekenstein,
  Phys.\ Rev.\ D {\bf 48}, 3641 (1993)
  [gr-qc/9211017].

\bibitem{Bettoni:2013diz} 
  D.~Bettoni and S.~Liberati,
  Phys.\ Rev.\ D {\bf 88}, 084020 (2013)
  [arXiv:1306.6724 [gr-qc]].

  
\bibitem{Gleyzes:2014rba} 
  J.~Gleyzes, D.~Langlois and F.~Vernizzi,
  Int.\ J.\ Mod.\ Phys.\ D {\bf 23}, no. 13, 1443010 (2015)
  [arXiv:1411.3712 [hep-th]].
  
   
\bibitem{Langlois:2017mxy} 
  D.~Langlois, M.~Mancarella, K.~Noui and F.~Vernizzi,
  JCAP {\bf 1705}, no. 05, 033 (2017)
  [arXiv:1703.03797 [hep-th]].
  
\bibitem{Kobayashi:2014ida} 
  T.~Kobayashi, Y.~Watanabe and D.~Yamauchi,
  Phys.\ Rev.\ D {\bf 91}, no. 6, 064013 (2015)
  [arXiv:1411.4130 [gr-qc]].
  
\bibitem{Koyama:2015oma} 
  K.~Koyama and J.~Sakstein,
  Phys.\ Rev.\ D {\bf 91}, 124066 (2015)
  [arXiv:1502.06872 [astro-ph.CO]].
  
\bibitem{Saito:2015fza} 
  R.~Saito, D.~Yamauchi, S.~Mizuno, J.~Gleyzes and D.~Langlois,
  JCAP {\bf 1506}, 008 (2015)
  [arXiv:1503.01448 [gr-qc]].
 
\bibitem{Babichev:2016jom} 
  E.~Babichev, K.~Koyama, D.~Langlois, R.~Saito and J.~Sakstein,
  Class.\ Quant.\ Grav.\  {\bf 33}, no. 23, 235014 (2016)
  [arXiv:1606.06627 [gr-qc]].
  
\bibitem{Sakstein:2016oel} 
  J.~Sakstein, E.~Babichev, K.~Koyama, D.~Langlois and R.~Saito,
  Phys.\ Rev.\ D {\bf 95}, no. 6, 064013 (2017)
  [arXiv:1612.04263 [gr-qc]].

  
\end{thebibliography}
\end{document}